\documentclass[twocolumn,prb,showpacs,superscriptaddress]{revtex4}
\usepackage{graphicx}
\usepackage{dcolumn}
\usepackage{bm}

\begin{document}

\title{Effect of transition-metal substitution in iron-based superconductors}

\author{S. L. Liu}
\affiliation{College of Science, Nanjing
University of Posts and Telecommunications,\\
Nanjing 210003, China}

\author{Tao Zhou}
\affiliation{College of Science, Nanjing University of Aeronautics
and Astronautics, \\
Nanjing 210016, China}

\date{\today}

\begin{abstract}

We study theoretically the effect of transition-metal (TM) substitution in iron-based superconductors through treating all of the TM ions as randomly distributed impurities. The extra
electrons from TM elements are localized at the impurity sites. In
the mean time the chemical potential shifts upon substitution. The
phase diagram is mapped out and it seems that the TM elements can
act as effective dopants. The local density of states (LDOS) is
calculated and the bottom becomes V-shaped as the impurity
concentration increases. The LDOS at the Fermi energy
$\rho(\omega=0)$ is finite and reaches the minimum at the optimal
doping level. Our results are in good agreement with the scanning
tunneling microscopy experiments.

\end{abstract}


\maketitle

The iron-based superconductors have been studied intensively since
their discovery~\cite{Kamihara2008}. It is widely believed that
the main physics in this family of materials is with the
iron-arsenic planes. Superconductivity can be realized by doping either
holes or
electrons into the system.

The transition-metal (TM) elements (such as Co or Ni) are widely used to
achieve superconductivity by substituting the iron
ions~\cite{Sefat2008,Chu2009,Fang2009}. However, this substitution
is significantly different from other doped materials, namely, the
TM ions enter the conducting planes and may also act as the
impurities. The competition of the doping effect and impurity
effect in the TM-doped materials is of great interest. The
scattering effect induced by the impurity has been studied
intensively. It has been proposed to account for many unusual
physical
properties~\cite{Zhang2009,Bang2009,Parker2008,Vorontsov2009,Tsai2009,Plamadeala2010,Zhou2011,zhou,jiang,kem}.
Since it is expected that all of the TM-ions enter the conducting
planes, the impurity concentration is expected to equal to the doping
density. While so far little attention has been paid to this issue
when studying the impurity effect, which may account for some
unusual experimental observations of the TM-doped compound.


Recently, the
TM substitution effect has attracted broad interest~\cite{Wadati2010,naka,Konbu2011,Berlijn2011,Bittar2011}. For Co(Ni) substitutes, it was reported
that the extra electrons are concentrated
 at the Co or Ni sites based on the first-principle
density functional method~\cite{Wadati2010,naka,Konbu2011,Berlijn2011}.
 Very recently, based on the X-ray absorption experiment it was also indicated
 that the electronic occupations for iron sites keep constant as Cobalt density changes in the
BaFe$_{2-2x}$Co$_{2x}$As$_2$ compound~\cite{Bittar2011}. These
results lead to the fundamental question, namely, whether the TM
substitutes provide doping carriers to the system. On the other hand,
the numerical calculation based on the first principle calculation
revealed that the Fermi energy shifts upon TM substitution although
the extra electrons are localized around the impurity
site~\cite{naka,Konbu2011,Berlijn2011}. The angle-resolved
photoemission spectroscopy (ARPES) experiments on
BaFe$_{2-2x}$Co$_{2x}$As$_2$~\cite{Sekiba2009,liu} have indicated
the evolution of the Fermi surface with Co-substitution, which
also seems to propose that the Cobalt atoms could be treated as
dopants. Thus at this stage the TM substitution effect is non-trivial and rich in physics. Studying this issue theoretically may provide
insightful hints to clarify the mechanism of superconductivity.

 In this Letter, we study the Cobalt doped material and treat
 all of the Cobalt ions as the randomly distributed impurities.
 The order parameters and the Fermi energy are obtained self-consistently
based on the two-orbital model and the Bogoliubov-de-Gennes (BdG)
equations. The extra electrons are totally located at the
impurity sites and the electron filling on Fe site keeps the same upon
substitution. In the parent compound, the
spin-density-wave (SDW) order is revealed due to the Fermi surface
nesting. A rigid shift of the Fermi energy occurs upon
substitution, which would break the Fermi surface nesting and
suppress the SDW order. As a result, the superconducting (SC) order shows
up. These results are qualitatively the same with the previous
ones in the clean system~\cite{Zhou2010}. Thus we suggest that the
Cobalt substitutes should be treated as the effective dopants in
spite that the extra electrons are localized. The local density of
states (LDOS) is also studied and we propose that the disordered
impurities are necessary to elucidate some striking features revealed by the scanning tunneling microscopy (STM) experiments.

We start from a two-orbital model including the hopping elements, pairing term, on-site interactions and impurity part~\cite{Zhang2009,Zhou2010,Zhou2011}, expressed by
\begin{equation}\label{1}
    H=H_{t}+H_{\Delta}+H_{\mathrm{int}}+H_{\mathrm{imp}}.
\end{equation}
The hopping term $H_{t}$ can be expressed by
\begin{eqnarray}\label{2}
    H_{t}=-\sum_{i\mu{j\nu}\sigma}(t_{i\mu{j}\nu}c_{i\mu\sigma}^{\dagger}c_{j\nu\sigma}+\mathrm{H.c.})-t_{0}\sum_{i\mu\sigma}c_{i\mu\sigma}^{\dagger}c_{i\mu\sigma},
\end{eqnarray}
where $i$,$j$ are the site indices and $\mu,\nu=1,2$ are the
orbital indices, and $t_{0}$ is the chemical potential. The second
term is the paring term, which reads
\begin{eqnarray}\label{3}
    H_{\Delta}=\sum_{ij}(\Delta_{i\mu{j}\nu}c_{i\mu\sigma}^{\dagger}c_{j\nu\bar{\sigma}}^{\dagger}+\mathrm{H.c.}).
\end{eqnarray}
The on-site interaction term $H_{\mathrm{int}}$ can be written
as\cite{Zhou2010,Zhou2011,zhou,Jiang2009}
\begin{eqnarray}\label{4}
    H_{\mathrm{int}}=&&U\sum_{i\mu\sigma\neq\bar{\sigma}}\langle{n_{i\mu\bar{\sigma}}}\rangle{n_{i\mu\sigma}}+U'\sum_{i\mu\neq\nu,\sigma\neq\bar{\sigma}}\langle{n_{i\mu\bar{\sigma}}}\rangle{n_{i\nu\sigma}}
    \nonumber\\
    &&+(U'-J_{H})\sum_{i,\mu\neq\nu,\sigma}\langle{n_{i\mu\sigma}}\rangle{n_{i\nu\sigma}},
\end{eqnarray}
where $n_{i\mu\sigma}$ is the density operator at the site $i$
and orbital $\mu$. The inter-orbital onsite interaction $U^{\prime}$ is taken to be $U-2J_{H}$.

$H_{\mathrm{imp}}$ is the impurity part of the Hamiltonian. For Cobalt substitutes, this term should include the potential scattering and the magnetic part.
Here we consider the potential scattering at the Cobalt substitutes,
which should play dominant role based on the first-principle calculation~\cite{kem,Wadati2010,naka,Konbu2011,Berlijn2011,Bittar2011}. Then this term is written as,
\begin{eqnarray}\label{5}
    H_{\mathrm{imp}}=\sum_{i_{\mathrm{m}}\mu\sigma}V_{s}c_{i_{\mathrm{m}}\mu\sigma}^{\dagger}c_{i_{\mathrm{m}}\mu\sigma}.
\end{eqnarray}.

The Hamiltonian can be diagonalized by solving the BdG
equations self-consistently,
\begin{equation}\label{6}
\sum_j \sum_\nu\left( \begin{array}{cc}
 H_{i\mu{j}\nu\sigma} & \Delta_{i\mu j\nu}  \\
 \Delta^{*}_{i\mu j\nu} & -H^{*}_{i\mu{j}\nu\bar{\sigma}}
\end{array}
\right) \left( \begin{array}{c}
u^{n}_{j\nu\sigma}\\v^{n}_{j\nu\bar{\sigma}}
\end{array}
\right) =E_n \left( \begin{array}{c}
u^{n}_{i\mu\sigma}\\v^{n}_{i\mu{\bar{\sigma}}}
\end{array}
\right),
\end{equation}
where the Hamiltonian $H_{i\mu{j\nu}\sigma}$ is expressed by,
\begin{eqnarray}\label{7}
     H_{i\mu{j}\nu\sigma}=-t_{i\mu{j}\nu}+[U\langle{n_{i\mu\bar{\sigma}}}\rangle+(U-2J_{H})\langle{n_{i\bar{\mu}\bar{\sigma}}}\rangle
     \nonumber\\
     +(U-3J_{H})\langle{n_{i\bar{\mu}\sigma}}\rangle+\sum_mV_{s}\delta_{i,i_{m}}-t_{0}]\delta_{ij}\delta_{\mu\nu}.
\end{eqnarray}
The SC order parameter $\Delta_{i\mu{j}\nu}$ and the local
electron density $\langle{n_{i\mu}}\rangle$ are obtained
self-consistently,
\begin{equation}\label{8}
    \Delta_{i\mu{j}\nu}=\frac{V_{i\mu{j}\nu}}{4}\sum_{n}(u_{i\mu\uparrow}^{n}\upsilon_{j\nu\downarrow}^{n*}+u_{j\nu\uparrow}^{n}\upsilon_{i\mu\downarrow}^{n*})\mathrm{tanh}\bigg(\frac{E_{n}}{2k_{B}T}\bigg),
\end{equation}
\begin{equation}\label{9}
    \langle{n_{i\mu}}\rangle=\sum_{n}|u_{i\mu\uparrow}^{n}|^{2}f(E_{n})+\sum_{n}|\upsilon_{i\mu\downarrow}^{n}|^{2}[1-f(E_{n})].
\end{equation}
Here $f(x)$ is the Fermi distribution function and
$V_{i\mu{j}\nu}$ is the pairing strength.

The LDOS is expressed by
\begin{equation}\label{11}
    \rho_{i}(\omega)=\sum_{n\mu}[|u_{i\mu\sigma}^{n}|^{2}\delta(E_{n}-\omega)+|\upsilon_{i\mu\bar{\sigma}}^{n}|^{2}\delta(E_{n}+\omega)],
\end{equation}
where the delta function $\delta(x)$ has been approximated by
$\Gamma/\pi(x^{2}+\Gamma^{2})$ with the quasiparticle damping
$\Gamma=0.01$.

The hopping constants used in this two-orbital model
are expressed as\cite{Zhang2009}
\begin{equation}\label{12}
    t_{i\mu,i\pm\hat{\alpha}\mu}=t_{1}
    ~~~~~~~~\hat{\alpha}=\hat{x},\hat{y},
\end{equation}
\begin{equation}\label{13}
    t_{i\mu,i\pm(\hat{x}+\hat{y})\mu}=\frac{1+(-1)^{i}t_{2}}{2}+\frac{1-(-1)^{i}t_{3}}{2},
\end{equation}
\begin{equation}\label{14}
    t_{i\mu,i\pm(\hat{x}-\hat{y})\mu}=\frac{1+(-1)^{i}t_{3}}{2}+\frac{1-(-1)^{i}t_{2}}{2},
\end{equation}
\begin{equation}\label{15}
    t_{i\mu,i\pm\hat{x}\pm\hat{y}\mu}=t_{4}~~~~~\mu\neq\nu.
\end{equation}

In the following presented results, the hopping constants are chosen to be
$t_{1-4}=1.0,0.5,-1.8,0.03$. The chemical potential $t_{0}$ is
determined by the average electron filling per site $\langle
n\rangle$ with $\langle n\rangle=2+x$, where $x$ is the impurity
concentration. The on-site Coulombic interaction is taken as
$U=3.26$ and Hund coupling $J_{H}=1.4$. The pairing is chosen as
next-nearest-neighbor (NNN) intraorbital pairing with the pairing
strength $V=1.0$, which will reproduce the $s_{\pm}$ paring
symmetry\cite{Ding2008,Mazin2008,Yao2009,Wang2009}. The numerical
calculation is performed on a $24\times24$ lattice at the periodic
boundary conditions. To calculate the LDOS, a $40\times40$
supercell technique is used.

\begin{figure}
    \includegraphics[width=9cm]{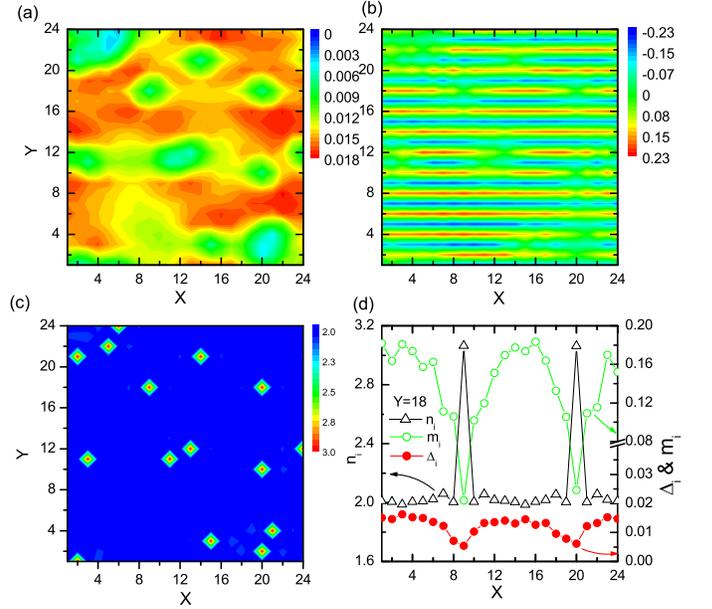}
\caption {(Color online) (a-c) The intensity plots of the SC order parameter, magnetic order and particle number with the impurity concentration $x=0.03$, respectively. (d) The two dimensional cut of the above parameters along $y=18$.}
\end{figure}

We plot the spatial distribution of the SC order parameter
$\Delta_{i}=\frac{1}{8}\sum_{\alpha\mu}\Delta_{i\mu,i+{\alpha}\mu}$,
the magnetic order
$[M_{i}=\frac{1}{4}\sum_{\mu}(n_{i\mu\uparrow}-n_{i\mu\downarrow})]$,
and the particle number $n_i=\sum_{\mu\sigma}n_{i\mu\sigma}$ with
the scattering potential $V_{s}=-3.5$ in Figs. 1(a)-1(c),
respectively. Fig. 1(d) displays the two dimensional cut of all of
the parameters. As seen, in presence of impurities, the amplitude
of the SC order parameter is not uniform. It is reduced at and around the impurity sites. The
magnetic spin order is antiferromagnetic along the $y$ direction
and ferromagnetic along the $x$ direction, which is consistent
with the $(\pi,0)$/$(\pi,\pi)$ SDW in the extended/reduced
Brillouin zone. This is qualitatively the same with the previous
theoretical calculation\cite{Zhou2010,Jiang2009}. At and around
the impurity sites, the magnetic order $M_{i}$ is suppressed. The
particle number is about $3.0$ at the impurity sites and recovers
to 2.0 when away from the impurity sites. This result indicates that
the electron filling for Fe sites does not change upon Cobalt
substitution. This is consistent with the numerical results based
on the first principle
calculation~\cite{Wadati2010,naka,Konbu2011,Berlijn2011}.

The average electron density per Fe site $n_{\mathrm{Fe}}$ and the
chemical potential $t_0$ as a function of the impurity
concentration $x$ are plotted in Fig. 2(a). As shown, the electron
density is almost unchanged ($\approx2.0$) for all of the impurity
concentrations we consider, indicating that the extra electrons of
impurity atoms are indeed localized around the impurity sites.
This is consistent with the first principle calculation
\cite{Wadati2010,naka,Konbu2011,Berlijn2011}. In the mean time,
the chemical potential increases  monotonically with increasing
substitution level $x$. This result confirms the numerical results
based on the first principle
calcuation\cite{naka,Konbu2011,Berlijn2011} and is consistent with
the ARPES experiments~\cite{Sekiba2009,liu}.

\begin{figure}
    \includegraphics[width=7cm]{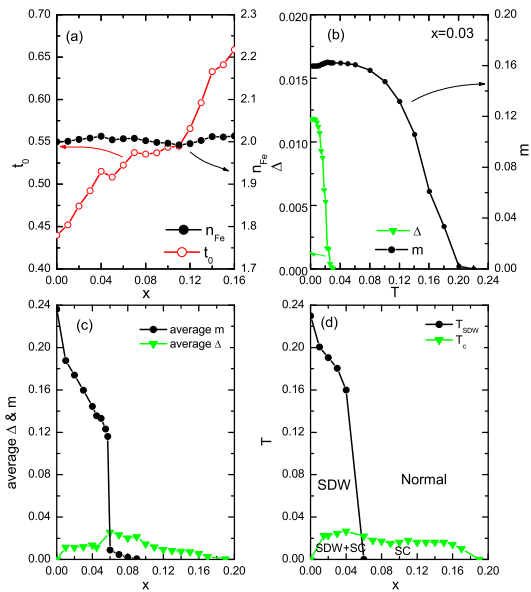}
\caption {(Color online) (a) The
chemical potential $t_{0}$ and the average particle number per Fe
ion $n_{\mathrm{Fe}}$ as a function of the Cobalt concentration $x$. (b) The average magnitudes of the SC order
parameter $\Delta$ and magnetic order $M$ as a function of the
temperature with the impurity concentration $x=0.03$. (c) The average
amplitudes of the SC order parameter $\Delta$ and the magnetic
order $M$ as a function of $x$. (d) The
calculated phase diagram.}
\end{figure}

The average amplitudes of the SC order and magnetic order as
a function of the temperature and impurity concentration $x$ are shown in Figs. 2(b) and 2(c),
respectively. For the fixed impurity concentration $x=0.03$, as seen in Fig. 2(b), both the magnetic order and the SC order
decrease as the temperature increases and two transition
temperatures are revealed. At zero temperature, as seen in Fig. 2(c), the
 magnetic order decreases monotonically  with
the increasing $x$ and vanishes around $x\simeq0.06$. The SC order
increases as $x$ increases in the low substitution region and
reaches its maximum at $x=0.06$. Then it decreases with further
increasing $x$. The calculated phase diagram is plotted in Fig. 2(d). It can be seen that the magnetic order and SC order coexist
in the low impurity concentration region. The magnetic order
decreases abruptly at $x=0.06$, corresponding to the quantum critical point at this concentration. This is well consistent with the experimental results on
BaFe$_{2-2x}$Co$_{2x}$As$_2$~\cite{lap,les}. The SC order appears as the magnetic order is
suppressed and the SC order reaches the
maximum as the magnetic order disappears. All of the obtained
results are qualitatively consistent with the previous calculation
by merely taking into account the doping effect~\cite{Zhou2010}.

\begin{figure}
    \includegraphics[width=7cm]{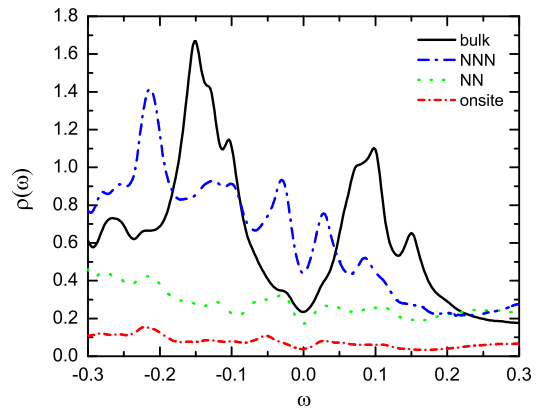}
\caption {(Color online) The energy dependent LDOS on the impurity
site, NN site and NNN site of the impurity, respectively, with the impurity concentration
$x=0.12$. The solid line represents the LDOS of bulk site which is far away from any impurity.}
\end{figure}

We have demonstrated that the Fermi energy increases as impurity concentration $x$ increases. The phase diagram and the order parameters as a function
of $x$ also indicate that the impurities are indeed act as the effective "dopant", in spite that the electron densities for Fe sites do not change.
While the fundamental question still remains about how the TM substitution controls the phase diagram.
Actually, for the TM-doped materials, the competition of the SDW and SC orders should play the essential role.
 The SDW instability originates from the Fermi surface nesting~\cite{Zhou2010}. It reaches the maximum value for the parent compound due to the
perfect nesting at zero doping.
The SC order is induced by the spin fluctuation~\cite{Mazin2008,Yao2009} and also relates to the Fermi surface nesting.
While for the parent compound the SC order could not survive due to the suppression effect by the SDW order. Both the impurity effect and the shifts of the Fermi energy would suppress the SDW order and then the SC order shows up upon substitution. Actually, here the disorder and the impurities play the important role in this system and to some extent they induce and enhance the SC order through suppressing the SDW order.

We now turn to discuss the LDOS spectra. Shown in Fig. 3 is the
LDOS spectra near an impurity with the impurity concentration
$x=0.12$. As seen, the spectra are suppressed at the impurity site
and the nearest-neighbor (NN) sites. Two in-gap peaks are
clearly seen on the NNN site to the impurity. These results are similar to the
previous theoretical results for single-impurity
effect~\cite{Zhou2011}.

\begin{figure}
    \includegraphics[width=7cm]{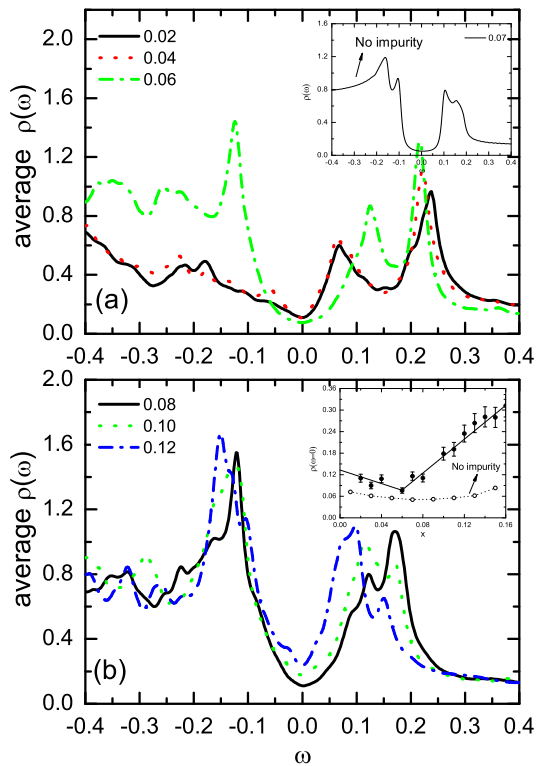}
\caption {(Color online) (a) The energy dependent LDOS of
$\mathrm{Ba(Fe_{1-x}Co_{x})_{2}As_{2}}$ superconductor in the low
substitution region. Inset:  The energy dependent LDOS with
doping $x=0.07$ without impurity. (b) Similar to (a)
but in the high substitution region. Inset: The solid circles are the LDOS at the Fermi energy $\rho(\omega=0)$ versus the impurity concentration $x$, with the solid lines being guides to the eye. The open circles and dashed line
are $\rho(\omega=0)$ versus the doping density without the impurity.}
\end{figure}

Now let us study the LDOS spectra in the bulk which are far away
from any impurity site. The average LDOS spectra of four sites are plotted
in Figs. 4(a) and 4(b), respectively. The bulk LDOS spectrum without the
impurity scattering at the doping level $x=0.07$ is plotted in the
inset of Fig. 4(a) for a comparison. As seen, for the clean system
the LDOS spectrum has a U-shaped bottom, consistent with
previous theoretical results~\cite{Zhou2010}. This is
understandable because the SC order parameter is nodeless around
the Fermi surface. While interestingly as the impurity
concentration increases the bottom becomes V-shaped. This is
consistent with the STM
experiments on $\mathrm{Ba(Fe_{1-x}Co_{x})_{2}As_{2}}$
superconductors\cite{Yin2009,Teague2011,Pan} and the previous
theoretical studies based on the T-matrix method~\cite{Bang2009}.
Another interesting feature of the spectrum is the finite value of
the LDOS at the Fermi energy [$\rho(\omega=0)]$. We plot the zero
energy LDOS as a function of the impurity concentration in the
inset of Fig. 4(b). $\rho(\omega=0)$ decreases with increasing $x$
and reaches the minimum at the optimal SC sample $x_m=0.06$. Then
it increases with further increasing $x$. As seen in the inset of
Fig. 4(b), $\rho(\omega=0)$ scales with $\mid x-x_m \mid$. This is
a striking feature and is well consistent with the STM
experiment\cite{Pan}. For a comparison we also plot
$\rho(\omega=0)$ as a function of doping density without
considering the impurity scattering. As seen, $\rho(\omega=0)$ is
very low and almost doping independent. Our results propose that
the impurity effect is essential when explaining some striking
features from STM experiments.

In summary, the TM substitution effect in iron-based
superconductors is studied by solving the BdG equations
self-consistently. We investigate the Cobalt substitutes and the Cobalt atoms are considered as randomly
distributed impurities with negative scattering potential. It is
shown that the extra electrons are localized around the impurity
sites, while the chemical potential increases with the increasing
impurity concentration. These results are consistent with the recent
first principle calculation and the ARPES experiments. The
obtained phase diagram is qualitatively consistent with the
previous reports of the BaFe$_{2-2x}$Co$_{2x}$As$_{2}$
superconductors. It is found that the energy dependent LDOS
spectrum is V-shaped, and the LDOS at the Fermi energy
$\rho(\omega=0)$ is finite and depends on the impurity
concentration. These features are consistent with the STM
experiments on BaFe$_{2-2x}$Co$_{2x}$As$_{2}$ material.

We thank S. H. Pan and Ang Li for useful discussion and showing us
their STM data before publication. This work is supported by the
NSFC under the Grant No. 11004105.

\end{document}